\documentclass[twocolumn,showpacs,superscriptaddress,email,floatfix,longbibliography,prl]{revtex4-2}

\usepackage[left=2cm,right=2cm,top=2cm,bottom=2cm]{geometry}
\usepackage{amsmath}
\usepackage{inputenc}
\usepackage{graphicx}
\usepackage{braket}
\usepackage{longtable}
\usepackage{bbold}
\usepackage{amsmath,amssymb,amsfonts,bbm,graphicx,hyperref,times,color}
\usepackage{multirow}
\usepackage{graphicx}
\usepackage[table,xcdraw]{xcolor}
\usepackage{natbib}
\usepackage{hyperref}
\usepackage{xr}

\begin{document}
\title{Dynamical phases and quantum correlations in an emitter-waveguide system with feedback}
\author{G. Buonaiuto}
\affiliation{Institut f\"ur Theoretische Physik, Universit\"at Tübingen, Auf der Morgenstelle 14, 72076 T\"ubingen, Germany}
\author{F. Carollo}
\affiliation{Institut f\"ur Theoretische Physik, Universit\"at Tübingen, Auf der Morgenstelle 14, 72076 T\"ubingen, Germany}
\author{B. Olmos}
\affiliation{Institut f\"ur Theoretische Physik, Universit\"at Tübingen, Auf der Morgenstelle 14, 72076 T\"ubingen, Germany}
\affiliation{School of Physics and Astronomy and Centre for the Mathematics and Theoretical Physics of Quantum Non-Equilibrium Systems, The University of Nottingham, Nottingham, NG7 2RD, United Kingdom}
\author{I. Lesanovsky}
\affiliation{Institut f\"ur Theoretische Physik, Universit\"at Tübingen, Auf der Morgenstelle 14, 72076 T\"ubingen, Germany}
\affiliation{School of Physics and Astronomy and Centre for the Mathematics and Theoretical Physics of Quantum Non-Equilibrium Systems, The University of Nottingham, Nottingham, NG7 2RD, United Kingdom}

\begin{abstract}
We investigate the creation and control of emergent collective behavior and quantum correlations using feedback in an emitter-waveguide system using a minimal model. Employing homodyne detection of photons emitted from a laser-driven emitter ensemble into the modes of a waveguide allows to generate intricate dynamical phases. In particular, we show the emergence of a  time-crystal phase, the transition to which is controlled by the feedback strength. Feedback enables furthermore the control of many-body quantum correlations, which become manifest in spin squeezing in the emitter ensemble. Developing a theory for the dynamics of fluctuation operators we discuss how the feedback strength controls the squeezing and investigate its temporal dynamics and dependence on system size. The largely analytical results allow to quantify spin squeezing and fluctuations in the limit of large number of emitters, revealing critical scaling of the squeezing close to the transition to the time-crystal. Our study corroborates the potential of integrated emitter-waveguide systems --- which feature highly controllable photon emission channels --- for the exploration of collective quantum phenomena and the generation of resources, such as squeezed states, for quantum enhanced metrology.
\end{abstract}

\maketitle

\textit{Introduction.} 
The development of techniques for the manipulation of matter with light has undergone rapid progress in the past decade \cite{mekhov2012,ritsch2013,hammerer2004,chou2005,browaeys2020,reiserer2015,duan2010,soare2014,hammerer2010}. This has enabled the creation of tailored quantum systems for the purpose of quantum simulation and information processing \cite{barrett2005,sangouard2011,georgescu2014}. It has also opened an avenue for the investigation of novel phases of matter and of the emergence of collective quantum behavior as it appears in the vicinity of a phase transition \cite{weimer2008,mann2018}. A paradigmatic example is the (open) Dicke model, which describes the interaction of an ensemble of atoms with a single-mode light field \cite{kirton2019}. It has received substantial attention in recent years, not only because of its fundamental theoretical interest, but also because it has been realized in various experimental platforms, such as atom-cavity setups \cite{baumann2010,zhiqiang2017}. An example of a novel many-body phase is a so-called time crystal. This is a phase of matter in which time-translation invariance is broken \cite{wilczek2012,shapere2012,sacha2017,else2020}. It has been theoretically predicted and analyzed in various scenarios, including disordered closed systems \cite{khemani2016,else2016,yao2017}, dissipative systems in continuous time \cite{iemini2018,buca2019a,hurtado-gutierrez2020} as well as periodic driven-dissipative systems \cite{sacha2015,lazarides2017,yao2020,gambetta2019a,gambetta2019b}. Moreover, this state of matter has been studied and characterized in a number of recent experiments \cite{choi2017,zhang2017,dogra2019,buca2019b}. 

In this work we discuss the many-body phases of a light-matter system that is composed of emitters coupled to light modes which propagate in a waveguide. This setup, which is subject matter of many current experimental studies \cite{corzo2016,meng2020}, has the appeal that the many-body dynamics can be reduced to a small set of degrees of freedom, and thus lends itself to a largely analytical treatment  \cite{lekien2014,ramos2014,pichler2015,kornovan2016,asenjo2017,lodahl2017,buonaiuto2019,jones2020,olmos2020,zhang2020}. We show that by implementing an instantaneous feedback protocol that relies on the measurement of a quadrature of guided light \cite{wiseman1994a,wiseman2009,jacobs2014,lammers2016,qi2016,nurdin2017,zhang2017,buonaiuto2020,ivanov2020,kroeger2020}, a rich variety of dynamical phases can be achieved, among them a continuous time crystal. For a large number of emitters, the dynamical and static phases of the system are described by a set of macroscopic spin variables whose expectation values obey a closed set of mean-field equations. We employ the theory of quantum fluctuations \cite{goderis1989a,goderis1989b,goderis1990,verbeure2010} to investigate the quantum correlations in the many-body state of the emitter ensemble. This mathematically rigorous approach, which becomes exact in the thermodynamic limit, has been used for isolated systems \cite{goderis1991a,lauwers2002,matsui2002,matsui2003,jakvsic2009,narnhofer2002,narnhofer2004,narnhofer2005}, to explore critical phenomena \cite{goderis1991b,verbeure1994,verbeure1995}, as well as in open quantum systems \cite{goderis1989c,benatti2017b,benatti2015,benatti2016b,benatti2018}, for instance to witness dissipative generation of entanglement in mesoscopic systems \cite{benatti2014,benatti2016a,benatti2017a}. Here, we use it to investigate spin squeezing of the emitter ensemble and to demonstrate critical power-law dynamics of quantum correlations at the boundary between two different non-equilibrium phases. Our results show that emitter-waveguide systems exhibit surprisingly complex emergent behavior and allow to realize collective states of matter on demand. This is not only of fundamental interest, but given the potential realizability of such systems as integrated devices, our work may stimulate applications in technological devices, e.g. for sensing and metrology, which are enhanced by quantum many-body effects.

\begin{figure}
    \includegraphics[width=\columnwidth]{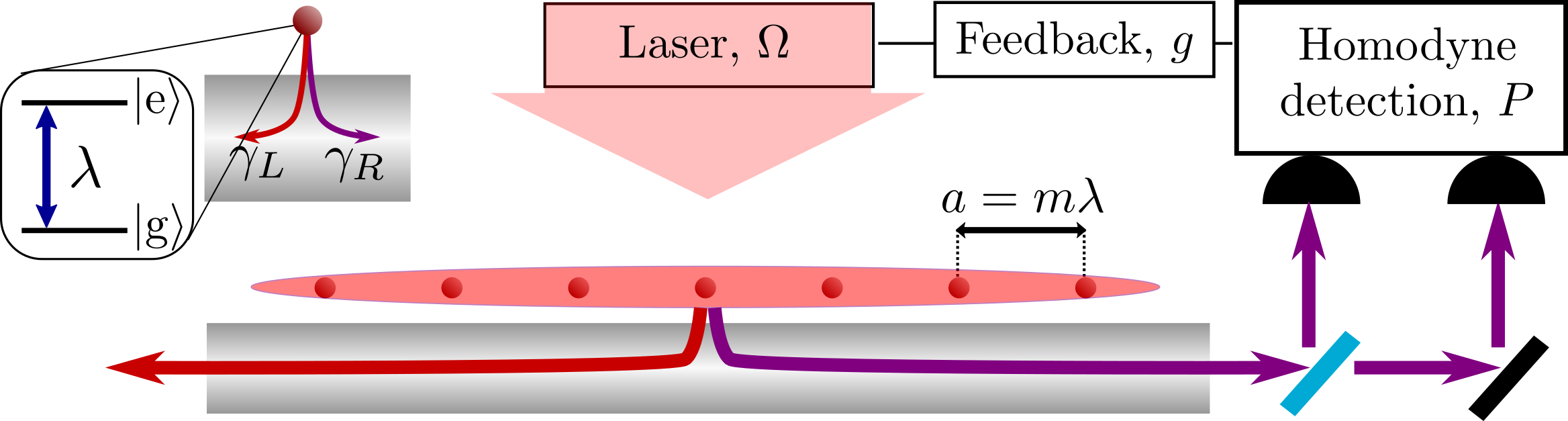}
    \caption{\textbf{Measurement-based feedback.} A chain of $N$ emitters (modelled as two-level systems) is placed in the vicinity of a waveguide and driven resonantly with Rabi frequency $\Omega$ by an external laser with wave vector perpendicular to the chain. Photons with wavelength $\lambda$ are emitted into the right- and left-propagating modes of the waveguide with equal emission rate $\gamma_R=\gamma_L$. When the separation between nearest neighbors in the chain, $a$, becomes commensurate with $\lambda$, the emission can be described as a collective process, where the chain of emitters is treated as a single macroscopic spin of length $N/2$. The measurement of the light emitted into the right-propagating mode via homodyne detection is fed back as a modulation of the laser field with strength $g$.}
    \label{fig:Setup}
\end{figure}

\textit{System and equations of motion.} The considered setup is sketched in Fig.~\ref{fig:Setup}. Here, a chain of $N$ emitters is placed in close vicinity to a waveguide, parallel to its longitudinal axis. An external laser field drives the emitters (modelled as two-level systems with ground and excited states $\left|\mathrm{g}\right>$ and $\left|\mathrm{e}\right>$, respectively) resonantly, with Rabi frequency $\Omega$. Photons are emitted from the chain into the \textit{guided} left- and right-propagating modes of the waveguide at a rate $\gamma_R$ and $\gamma_L$, respectively. We focus on the case in which the coupling between the emitters and the waveguide is non-chiral, i.e., the emission of photons into both guided modes occurs at the same rate, $\gamma_R=\gamma_L$. This can be in practice ensured by an appropriate choice of the laser polarization and transition dipole moment of each emitter. To allow an analytical treatment, our minimal model neglects photon losses into unguided radiation modes, c.f. Refs. \cite{ramos2014,pichler2015,zhang2020}. Choosing the laser wave vector perpendicular to the chain and the nearest neighbor distance $a$ between the emitters commensurate with the wavelength $\lambda$ of the emitted light, i.e. $a=m\lambda$ with $m=1,2,\dots$, in fact suppresses such unwanted decay channels \cite{buonaiuto2019,olmos2020}. Under these conditions and employing the Born-Markov approximation, the dynamics of the emitter-waveguide system is described by a particularly simple master equation:
\begin{equation}
\label{eq:ME}
\partial_t\rho=-\mathrm{i}2\Omega\left[J_x,\rho\right]+\gamma_R\mathcal{D}(J)\rho+\gamma_L\mathcal{D}(J)\rho.
\end{equation}
Here we have introduced the collective spin operators $J_{\alpha}=\frac{1}{2}\sum_{k=1}^N\sigma_\alpha^{(k)}$ with $\alpha=x,y,z$. The operators $\sigma_\alpha^{(k)}$ are Pauli matrices, e.g. $\sigma_x=\left|\mathrm{g}\right>\!\left<\mathrm{e}\right|+\left|\mathrm{e}\right>\!\left<\mathrm{g}\right|$, corresponding to the $k$-th emitter. Furthermore, $J=J_x-iJ_y$ is the lowering operator of the collective spin. The first term of Eq.~(\ref{eq:ME}) describes the resonant laser excitation of the emitters. The second and third term describe the incoherent emission of photons into the left- and right-propagating modes of the waveguide, through the dissipator $\mathcal{D}(J)\rho=J\rho J^\dagger-\frac{1}{2}\left\{J^\dagger J,\rho\right\}$.

The feedback scheme that we consider for the purpose of this work is based on the measurement of the phase quadrature $P(t)$ of the light emitted into the right-propagating mode. The measured quadrature of the photocurrent at each time determines the feedback, which consists of an instantaneous modulation of the Rabi frequency corresponding to the laser field driving the system. As shown in, e.g., Refs. \cite{lammers2016,buonaiuto2020}, the action of the feedback leads to the following modification to the master equation, 
\begin{eqnarray}
\label{eq:MEfeedback2}
    \partial_t\rho&=&-\mathrm{i}[2\Omega J_{x}-\frac{g\gamma}{2}\{J_{x},J_{y}\},\rho]\\
    &&+\frac{\gamma}{2}\mathcal{D}((2g+1) J_{x}-\mathrm{i}J_{y})\rho+\frac{\gamma}{2}\mathcal{D}(J_{x}-\mathrm{i}J_{y})\rho,\nonumber
\end{eqnarray}
where we have set $\gamma_R=\gamma_L=\gamma/2$. The dimensionless parameter $g$ represents the feedback strength. The action of the feedback is accounted for by a modified jump operator corresponding to emission into the right-propagating mode and a two-axis counter-twisting Hamiltonian term \cite{borregaard2017,kitagawa1993}. 

\textit{Dynamical and stationary phases.} In the following we study the thermodynamic limit, i.e. when $N\to\infty$. To this end we consider the dynamics of the expectation value of the magnetization operators $m_\alpha=J_\alpha/N$ with $\alpha=x,y,z$. In the thermodynamic limit, these operators permit a classical description of the system, since $\left[m_\alpha,m_\beta\right]=\mathrm{i}\varepsilon_{\alpha\beta\delta}m_\delta /N\approx 0$ as $N\to\infty$, and converge to their average value, $m_\alpha\to\langle m_\alpha\rangle$, for clustering states \cite{SM,lanford1969,verbeure2010}. This implies that expectation values involving them factorize, e.g., $\left<m_x m_z\right>\to \left<m_x\right>\left<m_z\right>$. Under the dynamics given by Eq.~\eqref{eq:MEfeedback2}, the exact time-evolution of these operators, in the large $N$ limit, is given by the mean-field equations \cite{SM,benatti2018}
\begin{eqnarray}
\label{eq:MF}
&&\partial_t\braket{m_{x}}=\Gamma\braket{m_{x}}\braket{m_{z}}\nonumber\\ &&
\partial_t\braket{m_{y}}=-2\Omega\braket{m_{z}}+\Gamma\kappa\braket{m_{y}}\braket{m_{z}}\\&&
\partial_t\braket{m_{z}}=2\Omega \braket{m_{y}}-\Gamma\braket{m_{x}}^{2}-\Gamma\kappa\braket{m_{y}}^{2}.\nonumber
\end{eqnarray}
Here we introduced the parameter $\kappa=2g+1$ and also the rescaled decay rate $\Gamma=\gamma N$, in order to ensure a well-defined thermodynamic limit \cite{kirton2019}. Eqs. (\ref{eq:MF}) conserve the length $j$ of the average magnetization vector $\vec{m}=(\braket{m_{x}},\braket{m_{y}},\braket{m_{z}})$, which, for the following considerations, we set to $\frac{1}{2}$: $j^2=\braket{m_{x}}^2+\braket{m_{y}}^2+\braket{m_{z}}^2=\frac{1}{4}$.

Let us first consider the particular case where $g=-\frac{1}{2}$, i.e. $\kappa=0$. Here, the mean-field equations of motion become particularly simple and one can, furthermore,  identify a second constant of motion, $\mathcal{C}_0=\braket{m_x} \exp{\left[\Gamma \braket{m_y}/(2\Omega)\right]}$, which simplifies the discussion considerably: throughout, we consider the initial conditions $\braket{m_x}(t=0)=\braket{m_y}(t=0)=0$ and $\braket{m_z}(t=0)=-\frac{1}{2}$, i.e.~the emitters are initially in the ground state $\bigotimes_{k=1}^N\left|\mathrm{g}\right>_k$, implying $\mathcal{C}_0=0$. Under this simplification the equations of motion of the average magnetization components $\braket{m_y}$ and $\braket{m_z}$ become that of a harmonic oscillator. It is apparent then that, for $\kappa=0$, the feedback eliminates the influence of dissipation and the magnetization undergoes persistent oscillations at frequency $2\Omega$. However, as we show below, this is not the case at the level of quantum fluctuations. These operators are indeed affected by dissipative effects which do not manifest in the dynamics of the magnetization operators. 

\begin{figure*}
\includegraphics[width=\textwidth]{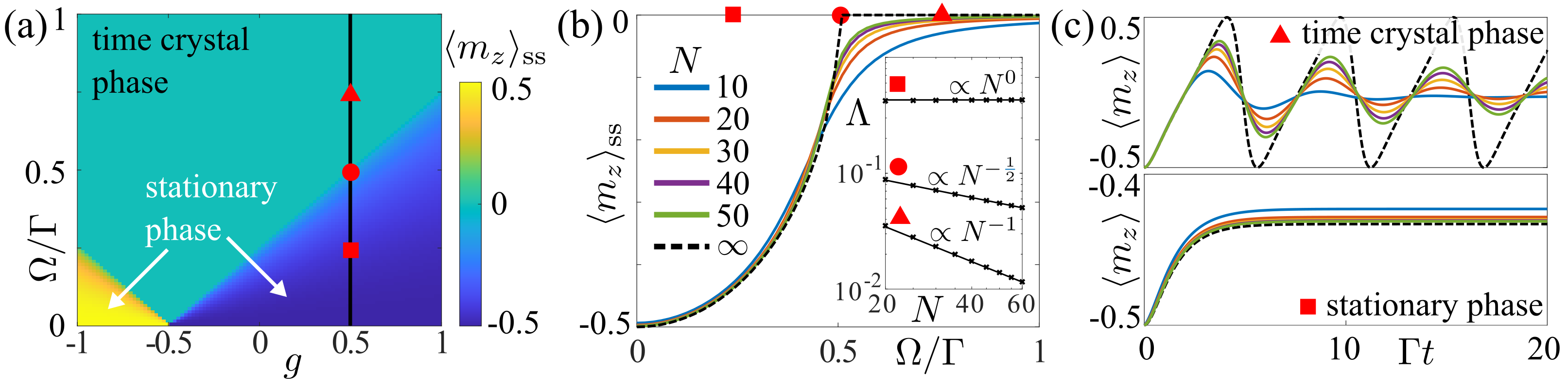}
\caption{\textbf{Phase diagram and mean-field vs. exact numerics.} (a) The mean-field diagram shows a time crystal phase and a stationary phase where $\braket{m_{z}}_\mathrm{ss}$ assumes a non-zero value. (b) Cut along the line $g=\frac{1}{2}$. The stationary value of $\braket{m_{z}}$ converges to the mean-field prediction (dashed line) as the number of emitters, $N$, increases. The transition to the time crystal phase takes place at $\Omega=\frac{1}{2}\Gamma$. As shown in the inset, at the critical point the spectral gap $\Lambda$ of the master operator (\ref{eq:MEfeedback2}) scales as $\propto N^{-\frac{1}{2}}$. In the time crystal phase the gap decreases as $N^{-1}$, while it is constant outside. (c) Time dependence of $\braket{m_{z}}$. Inside the time crystal phase $\braket{m_{z}}$ exhibits damped oscillations whose damping rate decreases proportional to $N^{-1}$, according to the scaling of the spectral gap. Outside the time crystal phase the system reaches a stationary state (stationary phase).}\label{fig:phase_diagram}
\end{figure*}

Away from this special point, the integration of the first two equations in (\ref{eq:MF}) yields the constant of motion
\begin{equation}
    \mathcal{C}_\kappa=\frac{\Gamma \braket{m_x}^{\kappa}}{\Gamma\kappa \braket{m_y}-2\Omega}.
\end{equation}
Again, we set $\mathcal{C}_\kappa=0$, which corresponds to initial conditions with $\braket{m_x}(t=0)=0$. Hence, the dynamics is reduced the two coupled equations
\begin{eqnarray}
\label{eq:MF2}
&&\partial_t\braket{m_{y}}=-2\Omega\braket{m_{z}}+\Gamma\kappa\braket{m_{y}}\braket{m_{z}}\\&&
\partial_t\braket{m_{z}}=2\Omega \braket{m_{y}}-\Gamma\kappa\braket{m_{y}}^{2},\nonumber
\end{eqnarray}
which can be further reduced to the second order differential equation
\begin{equation*}
    \partial^2_t\braket{m_{z}}=-4\Omega^2\left(1-\frac{\Gamma\kappa}{\Omega}\braket{m_{y}}\right)\left(1-\frac{\Gamma\kappa}{2\Omega}\braket{m_{y}}\right)\braket{m_{z}}. %\label{eq:newton_oscillator}
\end{equation*}
This is Newton's equation of motion for a particle with a non-linear restoring force, which supports oscillating solutions as long as the product of both brackets on the right hand side remains positive. There is clearly a parameter region where this condition can be met. This is delimited by the critical line $\Omega_c=\frac{1}{4}|\kappa|\Gamma_c$. Outside this oscillating regime the mean-field equations support a stationary solution, where the $z$-component of the mean-field vector follows
\begin{eqnarray}
\braket{m_{z}}_\mathrm{ss}=-\mathrm{sign}(\kappa)\sqrt{\frac{1}{4}-\frac{4\Omega^2}{\Gamma^2\kappa^2}}.
\end{eqnarray}

In Fig. \ref{fig:phase_diagram}(a) we show the corresponding phase diagram. One clearly recognizes the two boundaries which delimit the time crystal phase from a stationary phase where the $z$-component of the average magnetization assumes a finite constant value $\braket{m_{z}}_\mathrm{ss}$. Moreover, one also observes that for $g=-\frac{1}{2}$ ($\kappa=0$) the time crystal phase persists down to the limit of vanishing $\Omega$. In Fig.~\ref{fig:phase_diagram}(b) we show a cut through the phase diagram along $g=\frac{1}{2}$ in order to compare the result of the mean-field calculation with the exact numerics for finite $N$ obtained from Eq.~(\ref{eq:MEfeedback2}). One clearly sees that the numerical curves approach the mean-field prediction with increasing number of emitters $N$. This data confirms the position of the critical point as predicted by the mean-field equation (indicated by the red circle). Here the gap of the master operator (\ref{eq:MEfeedback2}) scales as $\Lambda\propto N^{-\frac{1}{2}}$. Fig.~\ref{fig:phase_diagram}(c) shows the time evolution of the $z$-component of the average magnetization vector. In the stationary phase this approaches quickly a stationary value which converges to the mean-field results as the number of emitters grows. In the time crystal phase the finite size simulations exhibit damped oscillations. As the number of emitters grows, the damping rate decreases and the mean-field solution predicts persistent oscillations in the large $N$ limit.

\begin{figure*}
\includegraphics[width=\textwidth]{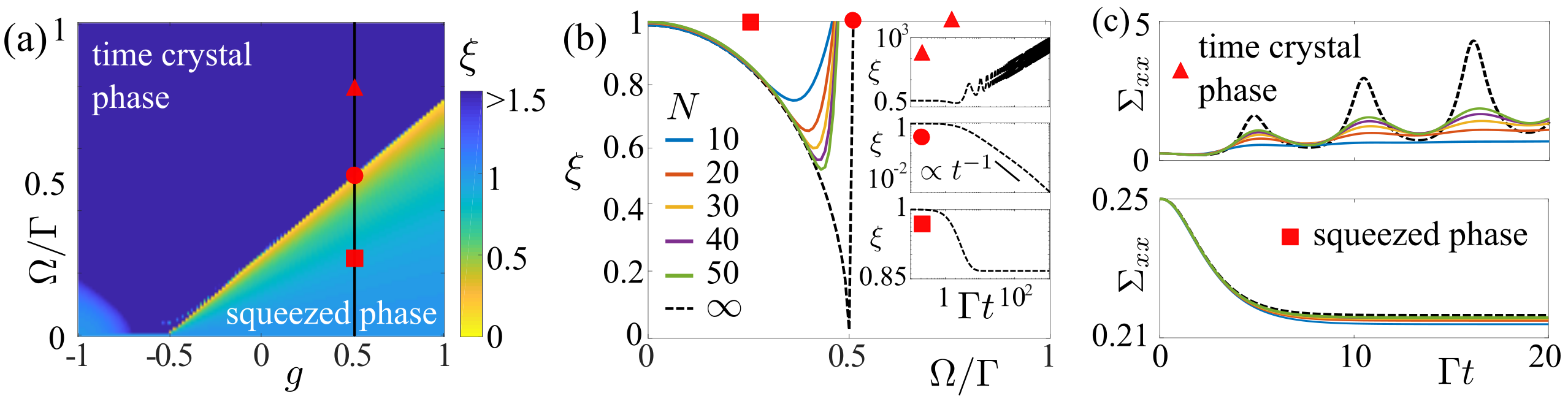}
\caption{\textbf{Spin squeezing.} (a) Long-time behavior of the spin squeezing $\xi$ in the thermodynamic limit [Eq.~\eqref{xi}], computed at time $\Gamma t=100$. In the time crystal phase, $\xi>1$ and tends to increase indefinitely [see the corresponding inset in panel (b)]. In the stationary phase, for $g>-\frac{1}{2}$, the emitter ensemble is spin squeezed. In this region, the smallest values of $\xi$ are reached at the transition line. In the stationary phase, for $g<-\frac{1}{2}$, $\xi$ approaches a finite value, which is larger than one. (b) The dashed line shows the behavior of $\xi$ in the thermodynamic limit [Eq.~\eqref{xi}] --- computed as in panel (a) --- as a function of $\Omega/\Gamma$, for $g=\frac{1}{2}$. For $\Omega/\Gamma>\frac{1}{2}$, the spin squeezing parameter increases indefinitely. Solid lines show a comparison with the spin squeezing parameter for finite-size systems [Eq.~\eqref{xi_N}], computed with the stationary state of Eq.~\eqref{eq:MEfeedback2}. The insets show the behavior of $\xi$ as a function of $\Gamma t$ for the different regimes, in the thermodynamic limit. For $\Omega/\Gamma<\frac{1}{2}$ the stationary state of the emitters becomes quantum correlated ($\xi<1$). For $\Omega/\Gamma>\frac{1}{2}$, $\xi$ grows indefinitely and features persistent oscillations. On the critical line, $\xi$ tends to zero with a power law decay $\xi\propto t^{-1}$. (c) Comparison between exact analytical results in the thermodynamic limit and finite-size numerics for the entry $\Sigma_{xx}$ of the covariance matrix. }\label{fig:spin-squeezing}
\end{figure*}

\textit{Fluctuations and quantum correlations}. We now turn to the analysis of quantum correlations. In particular, we seek to understand whether the feedback can generate spin squeezing, which is a measure of quantum correlations between the emitters and can quantify entanglement in many-body systems \cite{zhang2014,chen2016}. Squeezed spin states may find applications in metrological protocols, where they are used to enhance the precision of measurements beyond the standard quantum limit \cite{hosten2016}. Recently, spin squeezing in the steady state of dissipative systems has been explored in cavity QED setups \cite{masson2019} and in the non-equilibrium dynamics of an ensemble of superconduncting qubits \cite{xu2020}. In order to study quantum correlations within the emitter ensemble we will exploit the theory of quantum fluctuation operators \cite{goderis1989a,goderis1989b,goderis1990,verbeure2010}, applied to open quantum systems \cite{benatti2014,benatti2015,benatti2016b,benatti2016a,benatti2017a,benatti2017b,benatti2018}. For our model, this allows us to obtain rigorous analytical results in the thermodynamic limit.

The basic idea is to introduce operators describing fluctuations of the collective observables $J_\alpha$ around their average value $\langle J_\alpha\rangle$. Intuitively, these are given by  $F_\alpha=\frac{1}{\sqrt{N}}\left(J_\alpha-\langle J_\alpha\rangle\right)$, which are the quantum analogue of fluctuation variables, subject to central limit theorems \cite{goderis1989a}. To better understand their nature, we look at the commutators between fluctuations,
\begin{equation}
    \left[F_\alpha,F_\beta\right]=\mathrm{i}s_{\alpha\beta}\, , \quad s_{\alpha\beta}=\sum_\delta \epsilon_{\alpha\beta\delta}\, m_\delta\, , 
    \label{eq:F-comm}
\end{equation}
which are proportional to magnetization operators. As such, these commutators converge to scalar quantities, since $m_\delta\to\langle m_\delta\rangle$ for large $N$ \cite{SM,lanford1969,verbeure2010}, showing that fluctuation operators are bosonic operators in the thermodynamic limit. In fact, it is also possible to obtain the normal modes for quantum fluctuations, which behave as position and momentum operators. To do so, we rotate the coordinate system in such a way that the $z$-axis aligns with the direction $\vec{n}=\vec{m}/|\vec{m}|$ of the average magnetization vector. Here, we find that $\tilde{F}_z$ (the operator $F_z$ in the rotated frame) is a classical degree of freedom that commutes with the other fluctuations \cite{SM,benatti2016b,benatti2018}. For the directions perpendicular to $\vec{n}$, we can instead define ``position" and ``momentum" operators, $\hat{x}=\tilde{F}_x/\sqrt{j}$ and $\hat{p}=\tilde{F}_y/\sqrt{j}$, obeying canonical commutation relation $[\hat{x},\hat{p}]=\mathrm{i}$. 

To show that these bosonic fluctuation operators account for correlations between the emitters, we consider their covariance matrix. For the fluctuations $F_\alpha$, this is defined as
\begin{equation}
    \Sigma_{\alpha\beta}=\frac{1}{2}\langle \left\{F_\alpha ,F_\beta\right\}\rangle\! =\!\frac{1}{N}\!\left(\!\frac{1}{2}\langle \left\{J_\alpha ,  J_\beta\right\}\rangle \!-\!\langle J_\alpha\rangle \langle  J_\beta\rangle \!\right) . 
    \label{eq:F-cov}
\end{equation}
Contrary to what happens to magnetization operators --- which play the role, in this quantum context, of the sample mean variable of the law of large numbers --- correlations between fluctuations, as given by the quantities in Eq.~\eqref{eq:F-cov}, are not vanishing for $N\gg1$. Moreover, for sufficiently clustering states \cite{goderis1989a,goderis1989b,goderis1990,verbeure2010}, fluctuations possess a Gaussian distribution and the matrix elements $\Sigma_{\alpha\beta}$ are not divergent. In order to obtain the $2\times2$ covariance matrix of the fluctuations $\hat{x}$ and $\hat{p}$, we rotate $\Sigma$ [Eq.~(\ref{eq:F-cov})] into the new reference frame and extract the $2\times2$ principal minor rescaled by $j^{-1}$ \cite{SM}.  

The construction so far is associated with the quantum state of the emitter ensemble at a fixed time. In order to find the time-evolution of the canonical fluctuation operators $\hat{x}$ and $\hat{p}$ we need to consider the time-dependence of the matrices $s$ and $\Sigma$ [Eqs.~(\ref{eq:F-comm}) and (\ref{eq:F-cov})] in the large $N$ limit. These can be obtained from the master equation Eq.~\eqref{eq:MEfeedback2}, and by exploiting results from Refs.~\cite{benatti2016b,benatti2018}. As shown in the supplemental material \cite{SM}, ones finds that \begin{equation}
\Sigma(t)=X_{t,0}\Sigma(0)X_{t,0}^T-\int_0^t \mathrm{d}u\, X_{t,u} \, s(u) A s(u) \, X_{t,u}^T\, . 
\label{eq:Cov-t}
\end{equation}
Here $A=(\Gamma/2)\,\mathrm{diag}\left[1+\kappa^2,2,0\right]$, $X_{t,0}$ is the time-ordered exponential of the matrix
\begin{equation}
G\!=\!\left(\!
\begin{array}{ccc}
 \Gamma\langle m_z\rangle & 0 &  \Gamma  \langle m_x\rangle \\
 0 & \kappa\Gamma \langle m_z\rangle &  \kappa\Gamma \langle m_y\rangle  -2 \Omega  \\
 -2\Gamma \langle m_x\rangle    & 2 \Omega -2\kappa \Gamma \langle m_y\rangle  & 0 \\
\end{array}
\!\right)
\label{eq:G}
\end{equation}
and $X_{t,u}=X_{t,0}X_{u,0}^{-1}$. It is noteworthy that the dynamics of the fluctuations, as described by Eq. (\ref{eq:Cov-t}), cannot be obtained by linearizing the mean-field equations \eqref{eq:MF}. This means that fluctuations are affected by a dynamical building-up of correlations which is not captured by magnetization operators \cite{benatti2016b}. Moreover, the shape of Eq. (\ref{eq:Cov-t}) shows that the dynamics preserves the Gaussian character of fluctuations \cite{benatti2018}.

We are now in position to analyze spin squeezing. Several measures exist \cite{wineland1992,chase2008,ma2011,gross2012} and all of these are based --- in analogy with bosonic squeezing --- on the Heisenberg uncertainty principle. Here we consider the quantity \cite{kitagawa1993,gross2012}
\begin{equation}
    \xi=2\min\left(\Delta^2 J_\perp\right)/(N j)\, ,
    \label{xi_N}
\end{equation}
where $\Delta^2 J_\perp$ denotes the variance of a collective spin operator, $J_\perp$, living in the plane orthogonal to the principal direction $\vec{n}$ of the spin magnetization and the minimization is taken over all possible $J_\perp$. In the large $N$ limit, this measure converges to the bosonic squeezing parameter computed for the fluctuation operators \cite{SM}, and one finds that 
\begin{equation}
    \lim_{N\to\infty}\xi=2\min\left(\lambda_1,\lambda_2\right)\, ,
    \label{xi}
\end{equation} 
where $\lambda_i$ are the eigenvalues of the bosonic covariance matrix of $\hat{x}$ and $\hat{p}$. The state is spin squeezed if $\xi<1$. 
In Fig.~\ref{fig:spin-squeezing}(a-b), we show the behavior of the spin squeezing in the large $N$ limit. In the time-crystal phase this parameter does not converge to a stationary value [see also inset in Fig.~\ref{fig:spin-squeezing}(b) and Fig.~\ref{fig:spin-squeezing}(c), which displays one component of the covariance matrix]. On the other hand, in the stationary phase, for $g> -\frac{1}{2}$, the entries of the covariance matrix reach a stationary state [Fig.~\ref{fig:spin-squeezing}(c)]. Here, the squeezing parameter becomes smaller than $1$ [inset in Fig.~\ref{fig:spin-squeezing}(b)], which witnesses the presence of quantum correlations. Interestingly, at the critical line separating the time-crystal phase from the stationary one, spin squeezing displays a power-law behavior with time $\xi\propto t^{-1}$, suggesting a critical building-up of quantum correlations. Moreover, contrary to what one would expect from the phase diagram in Fig.~\ref{fig:phase_diagram}(a), the stationary phases to either side of $g=-\frac{1}{2}$ are different: for $g>-\frac{1}{2}$, fluctuations reach a stationary behavior and display non-trivial quantum correlations, while for $g<-\frac{1}{2}$, spin squeezing reaches a stationary value above $1$ and no stationary state for fluctuations exists. This qualitative change in behavior is rooted in the dependence of the structure of the matrix $G$ [cf. Eq. (\ref{eq:G})] on $\braket{m_z}$, which switches from positive to negative at $g=-\frac{1}{2}$.

\textit{Conclusions}. We have shown that feedback control of a coupled emitter-waveguide system offers the possibility to realize and manipulate dynamical many-body phases, such as a time crystal. Our minimal model allowed us to perform a largely analytical investigation which not only shed light on the behavior of the order parameter but also permitted the a rigorous analysis of fluctuations. This has shown that feedback can control spin squeezing of the quantum state of the emitters, which is particularly pronounced near the transition to the time crystal phase, where it exhibits critical scaling. In an actual experimental setting the finite number of emitters and also the existence of photon decay channels outside of the waveguide will impact on the long time behavior. The phases of the idealized setting discussed here will then be observed in the transient of the dynamics, i.e. they are expected to become meta-stable \cite{Kouzelis2020}. This has to be considered when seeking to design protocols that exploit, e.g., squeezing in technological applications such as quantum metrology.

\acknowledgments
\textit{Acknowledgements}. The research leading to these results has received funding from the European Union’s H2020 research and innovation programme [Grant Agreement No. 800942 (ErBeStA)]. We also acknowledge support from the “Wissenschaftler-R\"uckkehrprogramm GSO/CZS” of the Carl-Zeiss-Stiftung and the German Scholars Organization e.V., as well as through the Deutsche Forschungsgemeinsschaft (DFG, German Research Foundation) under Project No. 435696605, and under Germany’s Excellence Strategy - EXC No. 2064/1 - Project No.  390727645. BO was supported by the Royal Society and EPSRC [Grant No. DH130145].

%\bibliography{biblio}


\newpage
\mbox{~}
\newpage

\onecolumngrid

\renewcommand\thesection{S\arabic{section}}
\renewcommand\theequation{S\arabic{equation}}
\renewcommand\thefigure{S\arabic{figure}}
\setcounter{equation}{0}
\setcounter{figure}{0}

\begin{center}
{\Large SUPPLEMENTAL MATERIAL}
\end{center}
%\begin{center}
%\vspace{0.8cm}
%{\Large Dynamical phases and quantum correlations in an emitter-waveguide system with feedback}
%\end{center}
%\begin{center}
%G. Buonaiuto,$^{1}$ FC$^{1}$, BO$^{1,2}$, IL$^{1,2}$...
%\end{center}
%\begin{center}
%$^1${\em Institut f\"ur Theoretische Physik, Universit\"at %T\"ubingen,}\\
%{\em Auf der Morgenstelle 14, 72076 T\"ubingen, Germany}\\
%$^2${\em School of Physics and Astronomy and}\\
%{\em Centre for the Mathematics and Theoretical Physics of %Quantum Non-Equilibrium Systems,}\\
%{\em  University of Nottingham, Nottingham, NG7 2RD, UK}\\
%
%\end{center}
\section*{Recasting the dynamical generator}
\label{sec1}
In order to find the contribution to the dynamical equations of the different terms in the generator, we rewrite the latter in a convenient form. In addition, we work in the Heisenberg picture, where only operators are evolved. In particular, the Heisenberg representation of the generator appearing in equation \eqref{eq:MEfeedback2} can be written as the sum of four different terms 
\begin{equation}
\mathcal{L}[X]=\mathcal{H}_{\rm L}[X]+\mathcal{H}_{\rm C}[X]+\mathcal{A}[X]+\mathcal{B}[X]\, .
\label{gen-terms}
\end{equation}
The first contribution contains the local terms of the Hamiltonian and is given by 
\begin{equation}
\mathcal{H}_{\rm L}[X]=\mathrm{i}[H_{\rm L},X]\, , \qquad \mbox{ with } \qquad H_{\rm L}=\sum_{\mu=x,y,z}\omega_{\mu} J_\mu\, , \qquad \mbox{ and} \qquad \omega=(2\Omega,0,0)\, .
\label{gen1}
\end{equation}
The second contribution is still a coherent Hamiltonian one, and accounts for the coherent all-to-all interaction between emitters generated by the feedback. It can be written as 
$$
\mathcal{H}_{\rm C}[X]=\mathrm{i}[H_{\rm C},X]\, , \qquad \mbox{ with } \qquad H_{\rm C}=\frac{1}{N}\sum_{\mu,\nu} h_{\mu\nu}J_\mu J_\nu \, ,
$$
where $h$ is a $3\times3$ Hermitian matrix. In our specific setup, such a matrix  assumes the form
$$
h=-\frac{g\Gamma}{2}\begin{pmatrix}
0&1&0\\
1&0&0\\
0&0&0
\end{pmatrix}\, .
$$
The last two contributions are due to the dissipative part of the generator, which is split into two terms: 
\begin{equation}
\mathcal{A}[X]=\frac{1}{N}\sum_{\mu,\nu}\frac{A_{\mu\nu}}{2}\left[\left[J_\mu,X\right],J_\nu\right]\, ,\quad \mbox{ and } \quad \mathcal{B}[X]=\frac{\mathrm{i}}{N}\sum_{\mu,\nu}\frac{B_{\mu\nu}}{2}\left\{\left[J_\mu,X\right],J_\nu\right\}\, ,
\label{diss}
\end{equation}
where the matrices $A$ and $B$ are given by
$$
A=\frac{\Gamma}{2}\begin{pmatrix}
1+\kappa^2&0&0\\
0&2&0\\
0&0&0
\end{pmatrix}\qquad \mbox{ and }\qquad B=\frac{\Gamma}{2}\begin{pmatrix}
0&-(\kappa+1)&0\\
(\kappa+1)&0&0\\
0&0&0
\end{pmatrix}
$$
and $\kappa=2g+1$.

\section*{Mean-field dynamics}
In this section we give some details on the dynamics of the  magnetization operators for the emitters. Given the collective operators $J_\alpha$, these are defined as 
\begin{equation}
m_\alpha=\frac{J_\alpha}{N}\, ,
\end{equation}
and, due to the scaling $\frac{1}{N}$, such operators form a commutative algebra in the limit $N\to\infty$. This can be seen from the fact that $\|[m_\alpha,m_\beta]\|\propto \frac{1}{N}\to 0$, for $N\gg1$. 

In addition, the dynamics considered in the main text, as well as the initial state we focus on, are such that only very weak collective correlations between pairs of emitters \cite{benatti2016b,benatti2018} are present. These correlations are not inherited by the operators $m_\alpha$. As such, under the expectation $\langle \cdot\rangle$ over the state, these operators  $m_\alpha$ obey a sort of law of large numbers and converge, for $N\gg1$, to their expectation value 
\begin{equation}
 m_\alpha \longrightarrow \langle m_\alpha\rangle \, .
\label{mac-conv}
\end{equation}
This also means that the expectation of any product of magnetization operators converges to the product of the limiting operators, i.e.~the expectation value factorizes as
\begin{equation}
\langle m_\alpha m_\beta \rangle \longrightarrow \langle m_\alpha\rangle \langle  m_\beta\rangle \, .
\label{factorization}
\end{equation}

Our aim is thus to obtain the equations governing the dynamics of the quantities $\langle m_\alpha\rangle $ in the thermodynamic limit. These are the equations appearing in the main text in Eq.~\eqref{eq:MF}. To this end, we first compute the action of the generator on the collective operators $J_\alpha$ (using collective spin commutation relations):
$$
\mathcal{L}\left[J_\alpha\right]=-\sum_\delta\sum_{\mu}\omega_\mu \epsilon_{\mu\alpha\delta}J_\delta-\frac{1}{N}\sum_{\mu\nu}h_{\mu\nu}\left(\sum_\delta \epsilon_{\nu\alpha\delta}J_\mu J_\delta+\sum_\delta \epsilon_{\mu\alpha\delta}J_\delta  J_\nu\right)-\frac{1}{N}\sum_{\mu\nu}\frac{B_{\mu\nu}}{2}\sum_\delta \epsilon_{\mu\alpha\delta}\left\{ J_\delta,J_\nu\right\}+\mathcal{A}[J_\alpha]\, .
$$
We do not explicitly compute $\mathcal{A}[J_\alpha]$ since this is not relevant for the dynamics of $m_\alpha$. Using the above relation, we straightforwardly find the action of the generator on the  $m_\alpha$:
$$
\mathcal{L}\left[m_\alpha\right]\approx \sum_\delta D^{\rm L}_{\alpha\delta}\, m_\delta-\sum_{\mu\nu\delta}\left(h_{\mu\nu}+h^T_{\mu\nu}\right)\epsilon_{\nu\alpha\delta}\, m_\mu m_\delta-\sum_{\mu\nu\delta} B_{\mu\nu}\epsilon_{\mu\alpha\delta} m_\delta m_\nu\, .
$$
The approximate symbol appears because we have set to zero the  commutator of two average operators and because we have dropped the term $\mathcal{A}[J_\alpha]/N$ since both these terms converge to zero in norm, for large $N$. We  have further defined
$$
D_{\alpha\delta}^{\rm L}=-\sum_{\mu}\omega_\mu \epsilon_{\mu\alpha\delta}\, .
$$

In order to obtain the semi-classical equations of motion Eq.~\eqref{eq:MF}, which are exact in the thermodynamic limit, we consider the expectation of the above equation with respect to the quantum expectation $\langle \cdot\rangle$ recalling that, for such operators, expectation values factorize [see Eq.~\eqref{factorization}]. In this way, we find
\begin{equation}
\langle \dot{m}_\alpha\rangle =\sum_\delta D_{\alpha\delta}^{\rm L}\langle m_\delta\rangle-\sum_{\mu\nu\delta}\left(h_{\mu\nu}+h^T_{\mu\nu}\right)\epsilon_{\nu\alpha\delta}\, \langle m_\mu\rangle \langle m_\delta\rangle -\sum_{\mu\nu\delta} B_{\mu\nu}\epsilon_{\mu\alpha\delta} \, \langle m_\delta\rangle \langle  m_\nu\rangle \, ;
\label{mean-field}
\end{equation}
specializing the above equation for $\alpha=x,y,z$, we find the equations reported in Eq.~\eqref{eq:MF} in the main text.

\section*{Quantum Fluctuation Operators}
In order to quantify  quantum correlations in the emitter ensemble, as well as to derive their dynamical behavior,  we need to be able to control correlations between different collective spin operators, $J_\alpha$, with $\alpha=x,y,z$. We are interested in correlations of the form
$$
\Sigma_{\alpha\beta}=\frac{1}{N}\left(\frac{1}{2}\langle \left\{J_\alpha,J_\beta\right\}\rangle-\langle J_\alpha\rangle\langle J_\beta\rangle\right)\, ,
$$
where $\Sigma$ is nothing but a covariance matrix for collective spin operators. We want to look at the behavior of these correlations in the thermodynamic limit. To this end, it is convenient to exploit the formalism of quantum fluctuation operators \cite{goderis1989a,goderis1989b,goderis1990,verbeure2010}. 

For a given quantum expectation $\langle \cdot \rangle$, we can define the fluctuation operator $F_\alpha$ ($\alpha=x,y,z$) as
$$
F_\alpha=\frac{1}{\sqrt{N}}\left(J_\alpha-\langle J_\alpha\rangle\right)\, .
$$
This is essentially the collective operator $J_\alpha$ renormalized with respect to its expectation value in the state, and rescaled by a factor of $\frac{1}{\sqrt{N}}$. This operator is a quantum version of the fluctuation variable studied  in central limit theorems. Furthermore, note that
\begin{equation}
\langle F_\alpha\rangle=0\, ,\quad \mbox{ and }\qquad \Sigma_{\alpha\beta}=\frac{1}{2}\langle \left\{F_\alpha,F_\beta\right\}\rangle\, .
\label{expec-fluc}
\end{equation}
As shown in Refs.~\cite{goderis1989a,goderis1989b,goderis1990,verbeure2010}, under appropriate conditions on the quantum state defining the expectation $\langle \cdot \rangle$, which are always satisfied in our analysis, these fluctuation operators converge, for large $N$, to bosonic field operators. To show this, it is convenient to look at the commutation relations between fluctuation operators. These can be derived from the finite-$N$ commutation
$$
\left[F_\alpha,F_\beta\right]=\frac{\mathrm{i}}{N}\sum_\gamma \epsilon_{\alpha \beta \gamma}J_\gamma=\mathrm{i}s_{\alpha \beta}\, .
$$
In the above relations, the definition of $s_{\alpha\beta}$ makes apparent that such quantity has the scaling of a magnetization operator  and thus converges to a scalar multiple of the identity [see Eq.~\eqref{factorization}] under any expectation taken with the state $\langle \cdot \rangle$. In particular, the actual value to which this operator converges is 
$$
s_{\alpha\beta}\to \mathrm{i}\sum_\gamma \epsilon_{\alpha \beta \gamma}\langle m_\gamma\rangle \, .
$$
This shows that the commutator between two fluctuation operators, in the large $N$ limit, is actually a scalar multiple of the identity. Thus, fluctuation operators obey canonical commutation relations and can naturally be identified with bosonic operators. In our specific setting, we have three bosonic field operators whose commutation relations are encoded in the $3\times 3$ antisymmetric matrix
\begin{equation}
s=\begin{pmatrix}
0&\langle m_z\rangle &-\langle m_y\rangle \\
-\langle m_z\rangle &0&\langle m_x\rangle \\
\langle m_y\rangle &-\langle m_x\rangle &0\\
\end{pmatrix}\, .
\label{symplectic}
\end{equation}

In particular, for the setup considered here, we can find a mapping from these three generalized bosonic field operators to the standard bosonic position-like and momentum-like operators. Indeed, one can always find a rotation $R$ which brings $s$ into its canonical form 
$$
\tilde{s}=RsR^T=\begin{pmatrix}
0&j&0\\
-j&0&0\\
0&0&0\\
\end{pmatrix}\, ,
$$
with $R$ being a real unitary matrix, and $j=\sqrt{\langle m_x\rangle^2+\langle m_y\rangle^2+\langle m_z\rangle^2}$. Basically, the rotation simply consists in aligning the $z$-axis of the reference frame with the principal direction of the magnetization, $\vec{n}=j^{-1}\left(\braket{m_x},\braket{m_y},\braket{m_z}\right)$. This rotation $R$ thus identifies three new fluctuation operators $\tilde{F}_\alpha$ whose commutation relations are
$$
\left[\tilde{F}_\alpha,\tilde{F}_\beta\right]=i\tilde{s}_{\alpha\beta}\, ,
$$
and whose correlations are encoded in the appropriately rotated covariance matrix 
$$
\tilde{\Sigma}=R\Sigma R^T\, .
$$
We can then construct effective position and momentum operators 
$$
\hat{x}:=\frac{\tilde{F}_x}{\sqrt{j}}\, ,\qquad \qquad \hat{p}:=\frac{\tilde{F}_y}{\sqrt{j}}\, ,
$$
which are such that $[\hat{x},\hat{p}]=\mathrm{i}$. We also note that the operator $\tilde{F}_z$ plays the role of a scalar (classical) random variable since it commutes with the rest of the fluctuation algebra. For the relevant quantum degrees of freedom, $\hat{x}$ and $\hat{p}$, we can also compute the covariance matrix 
$$
\hat{\Sigma}=\begin{pmatrix}
\braket{ \hat{x}^2} & \frac{1}{2}\braket{\left\{\hat{x},\hat{p}\right\}}\\
\frac{1}{2}\braket{\left\{\hat{x},\hat{p}\right\}}& \braket{ \hat{p}^2}
\end{pmatrix}\, ,
$$
as the $2\times2$ principal minor of the matrix $\tilde{\Sigma}$, rescaled by the factor $j$. This covariance matrix contains the correlations between collective operators in the directions perpendicular to the principal magnetization vector $\vec{n}$. The smallest eigenvalue of $\hat{\Sigma}$ thus corresponds to the minimal variance that a collective operator, perpendicular to the direction $\vec{n}$,  can have, further rescaled by $j^{-1}$. When multiplied by $2$, the smallest eigenvalue of $\hat{\Sigma}$ is thus nothing but the spin squeezing parameter $\xi$, evaluated in the thermodynamic limit of a large number of emitters.

\section*{Time-evolution of the covariance matrix}
Now that we have understood the structure of fluctuation operators for a fixed state of the quantum system,  we need to be able to recover the covariance matrix $\Sigma$, at each time $t$. Essentially, we have to find how this matrix propagates in time according to the generator of the dynamics from a given initial condition. Before, though, note that, from Eq.~\eqref{symplectic}, it is clear that the dynamics of the anti-symmetric matrix $s$ is completely specified by the dynamics of the magnetization operators. 

To simplify the discussion, we introduce the matrix 
$$
C=\langle F_\alpha F_\beta\rangle\, ,
$$
whose connection with the  covariance matrix is explicitly
$$
\Sigma=\frac{1}{2}\left(C+C^T\right)\, .
$$
We start by taking the derivative of $C$. The time derivative of $F_\alpha$ is a scalar quantity since the derivative can only act on the state. This gives
$$
\frac{d}{dt}F_\alpha=-\frac{1}{\sqrt{N}}\frac{d}{dt}\langle J_\alpha\rangle\, ;
$$
recalling also  that $F_\alpha$ has a vanishing expectation on the state $\langle \cdot \rangle$, the time-derivative of the matrix $C$ is simply determined by the generator $\mathcal{L}$ as 
$$
\frac{d}{dt}C_{\alpha\beta}=\left\langle \mathcal{L}\left[F_\alpha F_\beta\right] \right\rangle\, .
$$
We now study the action of the different terms of the generator, appearing in Eq.~\eqref{gen-terms}, on the product of  two fluctuation operators. \\

First of all, we notice that 
\begin{equation}
\mathcal{H}_{\rm L}\left[F_\alpha F_\beta \right]=\mathcal{H}_{\rm L}\left[F_\alpha\right] F_\beta+F_\alpha\mathcal{H}_{\rm L}\left[F_\beta\right]=\sum_\gamma D^{\rm L}_{\alpha\gamma}\frac{J_\gamma}{\sqrt{N}}F_\beta+F_\alpha\sum_\gamma D^{\rm L}_{\beta\gamma}\frac{J_\gamma}{\sqrt{N}}\, .
\label{Loc1}
\end{equation}
Then, considering that the expectation value of a single fluctuation operator is zero under expectation over the state, we can freely subtract terms like $F_\alpha\sum_\gamma D^{\rm L}_{\beta\gamma}\frac{\langle J_\gamma\rangle}{\sqrt{N}}$ to the ones above, to reconstruct
$$
\left\langle \mathcal{H}_{\rm L}\left[F_\alpha F_\beta\right]\right\rangle=\sum_\gamma D^{\rm L}_{\alpha\gamma}C_{\gamma\beta}+\sum_{\gamma}D^{\rm L}_{\beta\gamma}C_{\alpha \gamma}\, .
$$
For the second term of the generator, the one with all-to-all interacting Hamiltonian, we also have 
\begin{equation}
\mathcal{H}_{\rm C}\left[F_\alpha F_\beta \right]=\mathcal{H}_{\rm C}\left[F_\alpha \right]F_\beta +F_\alpha \mathcal{H}_{\rm C}\left[F_\beta \right]\, .
\label{col1}
\end{equation}
We can thus focus on the action of $\mathcal{H}_{\rm C}$ on a single fluctuation operator. This gives 
\begin{equation}
\begin{split}
\mathcal{H}_{\rm C}\left[F_\alpha\right]=&\frac{\mathrm{i}}{N}\sum_{\mu\nu}h_{\mu\nu}J_\mu\left[ J_\nu,F_\alpha\right]+\frac{\mathrm{i}}{N}\sum_{\mu\nu}h_{\mu\nu}\left[J_\mu ,F_\alpha\right]J_\nu=\\
=&\mathrm{i}\sum_{\mu,\nu}h_{\mu\nu}F_\mu\left[F_\nu,F_\alpha\right]+\mathrm{i}\sum_{\mu,\nu}h_{\mu\nu}\left[F_\mu,F_\alpha\right]F_\nu\\
+&\mathrm{i}\sum_{\mu,\nu}h_{\mu\nu}\frac{\langle J_\mu\rangle}{N}\left[J_\nu,F_\alpha\right]+\mathrm{i}\sum_{\mu,\nu}h_{\mu\nu}\left[J_\mu,F_\alpha\right]\frac{\langle J_\nu\rangle}{N}\, .
\end{split}
\label{col2}
\end{equation}
To obtain the second equality, we have simply added and subtracted the proper expectation values, appearing in the last line of the above equation, in order to reconstruct fluctuation operators. We now divide the terms on the right hand side of the second equality into two contributions. We call  $\mathcal{H}_{\rm C}'$ the ones in the second line of the above equation and $\mathcal{H}_{\rm C}''$ those in the third line. 
The first  term can be understood by looking at commutation relation between fluctuation operators. Indeed, recalling the fact that $s_{\alpha\beta}=-s_{\beta\alpha}$ we can write
\begin{equation}
\begin{split}
\mathcal{H}_{\rm C}'\left[F_\alpha\right]&=-\sum_{\mu,\nu}h_{\mu\nu}F_\mu s_{\nu\alpha}-\sum_{\mu,\nu}h_{\mu\nu}s_{\mu\alpha}F_\nu=\\
&\approx\sum_{\gamma}Q^{\rm C}_{\alpha\gamma}F_{\gamma}\, ,
\end{split}
\label{col2}
\end{equation}
where we have 
$$
Q^{\rm C}=s\left(h^T+h\right)\, .
$$
We now come to the second term; this can be rewritten as  
\begin{equation}
\begin{split}
\mathcal{H}_{\rm C}''\left[F_\alpha\right]\approx \sum_{\gamma}D^{\rm C}_{\alpha\gamma}J_\gamma\, ,
\end{split}
\label{col3}
\end{equation}
where we have 
$$
D^{\rm C}_{\alpha\gamma}=-\sum_{\mu,\nu}\left[h_{\mu\nu}+h^T_{\mu\nu}\right]\epsilon_{\nu\alpha\gamma}\langle m_\mu\rangle \, .
$$
The approximate symbol in Eq.~\eqref{col3} is due to the fact that, in $D^{\rm C}$, we have considered $\langle m_{\mu}\rangle$ instead of $m_\mu$: this is only valid in the limit $N\to\infty$ [see also discussion of Eq.~\eqref{mac-conv}]. In addition, when considering the expectation over the state of the term in Eq.~\eqref{col1}, we have that we can safely add or substract a scalar quantity to the term $\mathcal{H}_{\rm C}''\left[F_\alpha\right]$ using the fact that, in any case, $F_\beta$ has zero expectation. Overall, we have found that
$$
\langle \mathcal{H}_{\rm C}\left[F_\alpha F_\beta \right]\rangle \approx \sum_\gamma\left(\left[Q^{\rm C}\right]_{\alpha\gamma}+\left[D^{\rm C}\right]_{\alpha\gamma}\right)C_{\gamma\beta}+\sum_\gamma\left(\left[Q^{\rm C}\right]_{\beta\gamma}+\left[D^{\rm C}\right]_{\beta\gamma}\right)C_{\alpha\gamma}\, .
$$
This concludes the contribution which comes from the Hamiltonian term of the generator. \\

We now turn to the dissipative contributions and start with the one encoded in $\mathcal{A}$. We have
\begin{equation}
\begin{split}
\left\langle \mathcal{A}\left[F_\alpha F_\beta \right]\right\rangle=\left\langle \frac{1}{N}\sum_{\mu,\nu}\frac{A_{\mu\nu}}{2}\left[\left[J_\mu,F_\alpha F_\beta \right],J_\nu\right]\right\rangle\, .
\end{split}
\end{equation}
To understand this contribution, we look at a single term in the above summation. This can be reduced to
\begin{equation}
\begin{split}
\left[\left[J_\mu,F_\alpha F_\beta \right],J_\nu\right]&=\left[\left[F_\mu ,F_\alpha \right],F_\nu \right]F_\beta +F_\alpha \left[\left[F_\mu ,F_\beta \right],F_\nu \right]\\
&+\left[F_\alpha ,F_\nu \right]\left[F_\mu ,F_\beta \right]+\left[F_\mu ,F_\alpha \right]\left[F_\beta ,F_\nu \right]\,  .
\end{split}
\label{diss1}
\end{equation}
The first two terms on the right hand side of the above equality are zero when taking the expectation over the state and in the thermodynamic limit. This can be understood as follows. All terms in the above equation actually consist of the product of two operators which have the scaling of magnetization operators. For the first two terms, one of the two is given by a fluctuation operator having a further scaling $\frac{1}{\sqrt{N}}$, which indeed transforms it into an operator with a scaling $\frac{1}{N}$. However, fluctuation operators are rescaled with respect to their average over the state so that, under any expectation over the state, the operator $F_\alpha/\sqrt{N}\to0$.

Concerning the remaining two terms [the ones in the second line of Eq.~\eqref{diss1}] we can use the fact that they are magnetization operators to argue that they converge to the product of their expectation. This leads to
$$
\left\langle \mathcal{A}\left[F_\alpha F_\beta \right]\right\rangle\approx -\left(sA\, s\right)_{\alpha\beta}\, .
$$

We are thus left with the second dissipative contribution, the one related to the matrix $B$, which acts on fluctuation operators as 
\begin{equation}
\begin{split}
 \mathcal{B}\left[F_\alpha F_\beta \right]&=\frac{\mathrm{i}}{N}\sum_{\mu\nu}\frac{B_{\mu\nu}}{2}\left\{\left[J_\mu,F_\alpha F_\beta \right],J_\nu\right\}=\\
 &=\mathrm{i}\sum_{\mu\nu}\frac{B_{\mu\nu}}{2}\left\{\left[F_\mu ,F_\alpha F_\beta \right],F_\nu \right\}+\mathrm{i}\sum_{\mu\nu}B_{\mu\nu}\frac{\langle J_\nu\rangle}{N}\left[J_\mu,F_\alpha F_\beta \right]\, .
\end{split}
\end{equation}
We divide the above term into two pieces. The first on the right hand side of the second equality we call it $\mathcal{B}'$ while the second $\mathcal{B}''$.
We focus on the first and, exploiting commutation relations of fluctuation operators, obtain
\begin{equation}
\begin{split}
 \mathcal{B}'\left[F_\alpha F_\beta \right]&=-\sum_{\mu\nu}\frac{B_{\mu\nu}}{2}\left\{s_{\mu\alpha} F_\beta ,F_\nu \right\}-\sum_{\mu\nu}\frac{B_{\mu\nu}}{2}\left\{F_\alpha s_{\mu\beta} ,F_\nu \right\}\, ;
\end{split}
\end{equation}
under expectation over the state this contributes with
\begin{equation}
\begin{split}
\left\langle \mathcal{B}'\left[F_\alpha F_\beta \right]\right\rangle&\approx -\sum_{\mu\nu}B_{\mu\nu}s_{\mu\alpha}\Sigma_{\nu\beta}-\sum_{\mu\nu}B_{\mu\nu}s_{\mu\beta}\Sigma_{\alpha\nu}\, .
\end{split}
\end{equation}
We are then left with the second part of this term. This gives
\begin{equation}
\begin{split}
\mathcal{B}''\left[F_\alpha F_\beta \right]&=\mathrm{i}F_\alpha \sum_{\mu\nu}B_{\mu\nu}\frac{\langle J_\nu\rangle}{N}\left[J_\mu,F_\beta \right]+\mathrm{i}\sum_{\mu\nu}B_{\mu\nu}\frac{\langle J_\nu\rangle}{N}\left[J_\mu,F_\alpha \right]F_\beta \\
&=-\sum_\gamma \left[\sum_{\mu\nu}B_{\mu\nu}\frac{\langle J_\nu\rangle}{N}\epsilon_{\mu\beta\gamma}\right]F_\alpha\frac{J_\gamma}{\sqrt{N}}-\sum_\gamma \left[\sum_{\mu\nu}B_{\mu\nu}\frac{\langle J_\nu\rangle}{N}\epsilon_{\mu\alpha\gamma}\right]\frac{J_\gamma}{\sqrt{N}}F_\beta\, .
\end{split}
\end{equation}
Under expectation we thus have
\begin{equation}
\begin{split}
\left\langle \mathcal{B}''\left[F_\alpha F_\beta \right]\right\rangle &\approx \sum_\gamma D^{\rm B}_{\alpha\gamma}C_{\gamma\beta}+\sum_\gamma D^{\rm B}_{\beta\gamma}C_{\alpha \gamma }
\end{split}
\end{equation}
with
$$
D^{\rm B}_{\alpha\gamma}=-\sum_{\mu\nu}B_{\mu\nu}\langle m_\nu\rangle \epsilon_{\mu \alpha\gamma}\, .
$$
Putting these results together, we obtain the exact differential equation for the evolution of $C$ which, in thermodynamic limit $N\to\infty$, becomes 
$$
\frac{d}{dt}C=-s A s +\left(D +Q^{\rm C} \right)C +C \left(D +Q^{\rm C} \right)^T+Q^{\rm B} \Sigma +\Sigma \left(Q^{\rm B} \right)^T
$$
where 
$$
D :=D^{\rm L} +D^{\rm C} +D^{\rm B} \, ,\qquad \mbox{ with } \qquad Q^{\rm B}=s B\, . 
$$
Finally, using the fact that $\Sigma =[C +C^T]/2$ we find \cite{benatti2016b,benatti2018} 
\begin{equation}
\frac{d}{dt}\Sigma =-s As +G \Sigma+\Sigma G^T \, ,
\label{diff-eq-cov}
\end{equation}
with 
$$
G =D +Q \, , \qquad \mbox{ and }\qquad Q =Q^{\rm C} +Q^{\rm B} 
$$
In our specific setting, the relevant matrices have the following form
\begin{equation}
D=\left(
\begin{array}{ccc}
 0 & 0 &  \Gamma  \langle m_x\rangle \\
 0 & 0 &  \kappa\Gamma \langle m_y\rangle  -2 \Omega  \\
 -\Gamma \langle m_x\rangle    & 2 \Omega -\kappa \Gamma \langle m_y\rangle  & 0 \\
\end{array}
\right)\qquad 
Q=\left(
\begin{array}{ccc}
 \Gamma\langle m_z\rangle  & 0  & 0 \\
 0   & \kappa\Gamma \langle m_z\rangle  & 0 \\
 -\Gamma\langle m_x\rangle  & -\kappa\Gamma\langle m_y\rangle & 0 \\
\end{array}
\right)\, .
\end{equation}
We recall here that all quantities $\langle m_\alpha\rangle$ are actually time-dependent and obey the system of differential equations \eqref{mean-field}. Whenever they appear in Eq.~\eqref{diff-eq-cov} they must be considered at the running time $t$. As such the actual solution of Eq.~\eqref{diff-eq-cov} is given by 
\begin{equation}
\Sigma(t)=X_{t,0}\Sigma(0)X_{t,0}^T-\int_0^t du X_{t,u}s(u)As(u)X_{t,u}^T\, ,
\label{Cov-t}
\end{equation}
where the matrix $s(t)$ is the matrix $s$ in Eq.~\eqref{symplectic} evaluated at time $t$ and $X_{t,u}$ is the time-order exponential of the matrix $G$, such that 
$$
\frac{d}{dt}X_{t,u}=G(t)X_{t,u}\, ,\qquad \mbox{and} \qquad \frac{d}{du}X_{t,u}=-X_{t,u}G(u)\, .
$$
\end{document}